\documentstyle[prb,aps,preprint,epsfig]{revtex}

\begin{document}
\title{Tunneling dynamics of side chains and defects in proteins, 
polymer glasses,
and OH-doped network glasses}

\author{Andreas Heuer$^{1)}$ and Peter Neu$^{2)}$}
\address{\it $^{1)}$Max-Planck Institut f\"ur Polymerforschung,
Ackermannweg 10, D-55128 Mainz, Germany}
\address{\it $^{2)}$Department of Chemistry,
 Massachusetts Institute of Technology, Cambridge, Ma 02139}

\maketitle

\thispagestyle{empty}

\begin{abstract} Simulations on a Lennard-Jones computer glass
are performed to study effects arising from defects in glasses at low
temperatures.
The numerical analysis reveals that already a low concentration of defects  may
dramatically
change the low temperature properties by giving rise to extrinsic
double-well potentials (DWP's).
The main characteristics of these extrinsic DWP's are
(i)  high barrier heights, (ii) high probability
that a defect is indeed connected with an extrinsic DWP,
(iii) highly localized dynamics around this defect, and (iv) smaller
deformation potential coupling
to phonons.
Designing an extension of the Standard Tunneling Model (STM) which parametrizes
 this picture and
comparing  with ultrasound experiments on the wet network glass
$a$-B$_2$O$_3$
shows that  effects of OH-impurities are accurately accounted for.
This model is then applied to organic polymer glasses and proteins.
It is suggested  that side groups may act similarly like  doped impurities 
inasmuch as
extrinsic DWP's  are induced, which possess a distribution of barriers peaked
around a high barrier height.
This compares with the structurlessly distributed barrier heights
of the intrinsic DWP's, which are associated
with the backbone dynamics. It is shown that this picture 
is consistent with elastic measurements on polymers, and
can explain anomalous nonlogarithmic line broadening recently observed 
in  hole  burning experiments in PMMA.
\end{abstract}

\setcounter{page}{0}

\section{Introduction}
Unlike simple systems such as atoms, nuclei or crystals, complex systems cannot
be entirely characterized  by the concept of a
ground state and elementary excitations: The ground state of systems such as
glasses, spin glasses or proteins is highly degenerate, and in place of energy
levels or
quasiparticles, we must speak in terms of a high-dimensional ``energy
landscape''
in configuration space. \cite{Stil,F2a,F1,F2b,F2,F3,TW,FS}
Basically the dynamics of the system is mapped on transitions between different
basins of the energy landscape.
Due to its complexity,  this landscape is very rough and
  only very qualitative features are known about it. Experiments on myoglobin
can be interpreted by saying that slightly different
structures---conformational
substate (CS),
separated by high barriers---are 
organized in  hierarchical tiers. \cite{F2a,F2,TW}
Glasslike behavior is also predicted for folded proteins (tertiary structure)
by both theory and experiment. \cite{F2b,Singh,Iben}
Somewhat closer information about the
multiple minimum  energy landscape also comes from computer simulations
on proteins, atomic clusters, and structural glasses. \cite{F2b,Berry,Heuer}

Whereas experiments at elevated temperatures (around the glass transition
temperature $T_g$) are sensitive to the overall topology of the energy
landscape,
experiments far below $T_g$ are sensitive to its fine structure.
In the framework of the standard tunneling model (STM)
this fine structure has been successfully characterized by
using a statistical approach. \cite{AHV,Phil}
The STM
explains many properties of glasses in the Kelvin regime.
It assumes that via local rearrangements the system can switch between
pairs of local energy minima.
Formally, this corresponds to the dynamics of a particle in a double-well
potential (DWP). At temperatures in the Kelvin regime the dynamics
has to be described in quantum-mechanical terms. The resulting
tunneling dynamics only involves the lowest two energy levels.
Hence the DWP's act as two-level systems (TLS's).
An important aspect of the STM is the assumption of a broad distribution
of parameters characterizing DWP's and hence TLS's. In particular,
this broad distribution  implies
that transition processes occur on a broad spectrum of time scales.

Most low temperature thermal, acoustic and optical properties of
glasses can be explained in this picture.  \cite{P} However, recent hole
burning
experiments on a polymer glass (PMMA) in Haarer's group report a systematic
disagreement with the STM. \cite{Haar1,Haar2,Haar3}
The authors  found a logarithmic time dependence with a crossover
to an algebraic behavior after ca. 3~h.
At the same  time equivalent measurements have been performed on a protein
(myoglobin)
in Friedrich's group \cite{FR3} with the result of almost
no hole broadening up to 3~h and an algebraic time dependence thereafter.
Very recently these results have been extended 
to longer waiting times.\cite{FR4}
Though the logarithmic behavior is in agreement with the STM, the algebraic
behavior is not.

Apparently unrelated anomalies can be observed in sound attenuation
experiments.
In  undiluted network glasses (SiO$_2$, GeO$_2$)
a single relaxation peak appears around 50 K which is consistent
with a broad distribution of DWP's. \cite{Hunk,N1,Topp} However,
an additional dilution with OH-impurities gives rise to the
 existence of a second relaxation peak at
higher temperatures. \cite{KK,N1}
 Specific mechanisms have been proposed giving
rise to DWP's connected with the OH-impurities. \cite{Phillips}

Interestingly, sound attenuation measurements on various  polymers
also show an additional strong  increase at temperatures 
above $\sim 30 - 50$~K.\cite{po1,po2,po4,po3,Pobell}
Different absorption peaks---labelled by Greek letters
 ($\alpha, \beta, \gamma, \delta$)
in appearance of decreasing temperature---have 
been observed in the temperature regime
between $\sim 30$~K and  room temperature.\cite{po1,po2} 
Usually, the low temperature peaks
($\delta,\gamma, \beta$) are attributed to side chain motions,
which  perform thermally activated 
reorientation processes even far below $T_g$.\cite{po1,po2,po4,Rohr} 

This  similarity  between   sound absorption  in OH-doped network
glasses and in polymer glasses may justify to think of side chain
as  ``defects'', because  both hydroxcyl groups and side groups give
rise to well-defined localized DWP's.
Thus one may ask the question whether the statistical approach of the STM still
holds  as soon as a significant fraction of DWP's
is related to the dynamics of some well-defined defects.

The conjecture we want to investigate in this paper is:
(1) the existence of minority components  of different structure
 (impurities as compared to the
network or side groups as compared to the main chain)
give rise to additional localized dynamical degrees of freedom;
(2) associated with these
degrees of freedom are induced or {\it extrinsic DWP's} which possess a
nonuniversal, i.e., different from the STM, distribution of parameters;
(3) these extrinsic DWP's cause a second absorption peak of
sound at higher temperatures and a nonlogarithmic hole broadening 
at ultralong time scales. In what follows both OH-impurities and side groups  
will  be referred to as defects.

Rather than just postulating the nonuniversal properties of defects,
we first show via computer simulations of a simple model glass that the
presence
of defects indeed has dramatic consequences for the distribution of DWP's.
We choose a binary Lennard-Jones (LJ) glass which
has  been  tested  in previous simulations and has proven to  confirm the
assumptions of the STM. \cite {HS}
The LJ parameters are chosen to represent NiP. 
To this LJ glass we add ``defects''
(minority component) in form of  a third  LJ particle. The size of the defect
will be controlled via its LJ parameters by a variable interaction length. The
main finding of these simulations is that,  in  case that the minority
component is somewhat
smaller than the original glass components, 
 they give rise to additional or extrinsic
DWP's with high energy barriers whereas in case of large defects
no extrinsic DWP's occur. The
existence of small  particles also has some impact on the {\it intrinsic}
DWP's which are connected with the majority component.
Although these results  might not
be representative for all glasses, one can speculate that as a general
result the distribution of DWP's in systems with defects can be
viewed to additionally comprise typical high barriers.
Based on this observation,  we design a model which extends the STM,
and compare to experiments on OH-diluted network glasses,  polymer glasses and proteins.

The outline of this paper is as follows:
In Sec. II,  we describe the simulation method and present the results;
in Sec. III,  an extension of the tunneling model
is proposed taking into account an additional
contribution of DWP's with high barrier heights;
in Sec. IV,  we compare with experiment, and, in Sec. V, we conclude.
Mathematical details are relegated to the Appendix in order not to interrupt
the main flow of the paper.

\section{Simulations}
\subsection{Model for the computer glass}

The model LJ-type glass  has been adapted
from the work of Weber and Stillinger \cite{Weber} and is devised
to represent the binary glass former NiP. The system
contains 80\% type 1 (Ni) and 20\% type 2 (P) particles.
The interaction is pairwise and reads
\begin{eqnarray}
V_{ij}  = \left\{ \begin{array}{lc}
A_{ij} [(\alpha_{ij} r/\sigma)^{-12} - 1] \exp[- \sigma/(a - \alpha_{ij}r)],
\qquad {\rm for}\quad    0 < \alpha_{ij} r \le a  \\
\hspace{2.5cm} 0\hspace{4.5cm}  \qquad {\rm else}
\end{array}\right.
\end{eqnarray}
with $a = 1.652 \,\sigma$ where $\sigma = 2.2$ {\AA} is the unit length.
Values for $A_{ij}$ and $\alpha_{ij}$ are
$A_{11} = 8200$ K, $\alpha_{11} = 1.0$ (Ni-Ni),
$A_{12} = 1.5\, A_{11}, \alpha_{12} = 1.05$ (Ni-P),
$A_{22} = 0.5\, A_{11}$, and $\alpha_{22} = 1.13$ (P-P).
The mass density $\rho_0$ is 8348 kg/m$^3$.
We have simulated systems with $N = 150$ particles
and use periodic boundary conditions. We have checked that within statistical
errors
a simulation with $N = 500$ particles yields identical results.
The equilibrium
distance between two particles of type $i$ and $j$
is given by  parameters $2^{1/6}\sigma / \alpha_{ij}$.
In extension to the above model we include a single
particle which we formally denote type 3 particle.
Its potential parameters are chosen as
$\alpha_{i3} = \alpha$ and $A_{i3} = A_{11}$. The
value of $\alpha$ determines the equilibrium distance
between this particle and the rest of the glass
and hence characterizes the effective size of this particle.
The simulations have been performed in the range
$ 0.7 \le \alpha \le 1.6$.
Of course, one expects that for $\alpha \sim 1$
this particle basically behaves like the rest of the glass for
which 96\% of all pair potentials are
characterized by $\alpha = 1$ or $\alpha = 1.05$.

\subsection{Simulation method}

In previous work Heuer and Silbey developed a systematic
search algorithm for DWP's. \cite{HS} Starting
from some $T=0$ configuration of a glass
the idea is to search the energy landscape around some local minimum
for a second local minimum. This pair of minima
can be viewed as a DWP. The full cooperativity of the transition
between both minima is taken into account.
In general all particles will be  involved in the transfer
between both minima.
Let $d_i$ denote the transfer distance of the $i$-th particle.
The total distance of both minima in configuration space
can then be expressed as
\begin{equation}
d^2 \equiv \sum_i \vec{d}_i^2.
\end{equation}
The algorithm takes advantages of the observation that
the existence of DWP's
is related to a somewhat localized motion of particles.
For different initial $T=0$ configurations
we analyzed the region around
the defect and checked whether we can find
a DWP close to this defect. This is achieved in
three steps.
In a first step we choose the defect particle as well as
its 15 nearest neighbors. Then in a second step we
minimize the potential energy via some simulated annealing
routine by moving the 16 selected particles such that
the distance in configuration space to the original
local minimum is $d_0 = 0.4\,\sigma$. By this condition
we avoid that the system just moves back to the original
 minimum. Finally, we relax the whole system to find
a new local minimum.  In case that close
 to the original minimum there exists a second
 minimum the algorithm will likely find this
new minimum. A careful
analysis for this choice of $d_0$ has shown that DWP's with
$d_0 < d < \sigma$ are systematically found by this
algorithm. Anyhow, the fraction of DWP's with $d < 0.4\, \sigma$
is very small.
In present work we restrict ourselves to DWP's with $d$
in this range.
For the binary LJ glass
it turned out that there exist one DWP per 100 particles.
Hence in a small region around the defect one expects
at most a single DWP.
Further details about this algorithm as well as
a discussion of its limitations can be found in Ref. \CITE{HS}.

We wish to distinguish the properties of extrinsic and intrinsic
DWP's. For this purpose,
we determine for all DWP's which particle moves most during
the transition between both local energy minima and denote this
distance by $d_s$. If this
particle is the defect we denote this DWP extrinsic,
otherwise intrinsic.
Finally the effective mass is defined via
\begin{equation}
p \equiv d^2 / d_s^2 \ge 1.
\end{equation}
The value of $p$ is a measure for the number of particles which are
basically involved in the formation of the DWP.  Finally,
the potential height of a DWP is found
by an explicit search for the saddle point.
The present analysis is based on the analysis of approximately 3000
independent initial configurations.

\subsection{Results of the simulations}
First we determine the probability that a DWP exists close to
the defect. In Fig. 1 the probability
is shown that the algorithm, described in Sec. IIA, finds  an
extrinsic DWP.
One can see that the number of extrinsic defects dramatically
increases for large values of $\alpha$. For the smallest
particles analyzed in our simulation runs nearly
80\% of all defects are connected with a DWP. Of course,
this number has to be viewed as a lower limit since it is
possible that our algorithm may miss some DWP's. However, this
already large number shows that the algorithm works
realibly. For $\alpha = 1$ the defect is basically identically
to the rest of the glass.
Hence the probability at $\alpha = 1$ has to be identical
to the probability (normalized per particle) to find a DWP in the
whole glass.
The present result
agrees with the already above-mentioned finding of Ref. \CITE{HS}
that on average one DWP exists per 100 particles.
For us it was very surprising that already for
$\alpha = 1.2$ a significant number of extrinsic DWP's
can be observed.
Hence only small differences between
defect and structural glass seem to strongly enhance the probability
for the formation of a DWP. For $\alpha = 0.7$ no extrinsic
DWP has been found. A huge particle is locked in a cage
formed by the surrounding small particles without a chance to
escape.

In Fig.~1 the probability to find intrinsic DWP's is
displayed, too. Originally, the simulation yields
the probability that there exists an intrinsic DWP's having a
significant spatial overlap with the initially selected particle cluster;
see the description of the algorithm above.
For $\alpha = 1$ this probability is larger than the probability
for the formation of extrinsic DWP's although no difference between
intrinsic and extrinsic DWP's should exist. The reason is trivially
related to the asymmetry in the definition of intrinsic and extrinsic
DWP's. For $\alpha = 1$ the ratio of
the number of intrinsic and extrinsic DWP's should be close to the
number of initially selected particles around the defect (here: 15 particles).
This unwanted statistical effect can be removed by scaling the probability
of formation of intrinsic DWP's such that for $\alpha = 1$ this
probability is equal to that of the formation of extrinsic DWP's.
This allows a direct comparison between intrinsic and extrinsic
DWP's. All data in Fig.~1 for the intrinsic DWP's are scaled in this way.
Interestingly, the number of
intrinsic DWP's close to the defect strongly depends on the
size of the defect. The number is largest for $\alpha \approx 1$, but
strongly decreases for larger or smaller defects.
We interpret this observation as follows: for $\alpha$ very different to
unity the defect
does not participate in the dynamics of the glass because it does not
move at all (small $\alpha$) or because it only moves by itself
(large $\alpha$). Hence from the viewpoint of the glass particles
there exists an ``alien element'' which forms a barrier
for the surrounding particles.
Thus close to the
defect the degrees of freedom and hence the chance to
form a DWP are significantly reduced. Only for $\alpha \approx 1$ this
effect is irrelevant since the defect may simply participate in
the dynamics.

In order to characterize the nature of the extrinsic DWP's somewhat
closer,  we plot the average effective mass $p$ of the DWP's
in dependence of $\alpha$; see Fig. 2. 
One observes  a transition from cooperative to strictly 
localized dynamics.
Already for $\alpha = 1.2$ the effective
mass of extrinsic DWP's starts to decrease. 
For $\alpha = 1.6$ the extrinsic
DWP's are related to motions of the defect alone. In the limit
of very small particles this result could have been expected.
Basically the defect jumps between different interstitial
positions formed by a fixed environment. 
It is likely
that for large $\alpha$ a defect has more than one direction in order to find
a second minimum. From the present simulations this suggestion cannot
be quantified.

In Fig. 3, we plot the distribution of barrier heights of intrinsic DWP's
with the presence of a close-by defect
as well as the distribution for extrinsic DWP's related to  $\alpha = 1.6$.
In both cases a broad distribution of barrier heights is observed
reflecting the statistical nature of the formation process of a DWP.
However, obviously the extrinsic defects in case of small defects
are significantly shifted to higher potential heights. This
is quantified in Fig. 4,
where  we plot the average barrier height $V$ in dependence of $\alpha$.
In agreement with Fig. 3,
one observes a strong increase of
the value of the average potential height with $\alpha$.
As already discussed in Ref. \CITE{HS},
the ability to cooperative motion, i.e.,  a large effective mass $p$,
tends to decrease the necessary energy changes along the way between both
minima of a DWP. In the opposite limit $p=1$ a single particle has to move
by itself, hence experiencing strong resistance by the environment leading
to a larger value of the barrier height.
This general statement is again confirmed  by the observed correlation
between the effective mass and the average barrier height; 
compare Figs. 2 and 4.

Finally, in Fig. 5, we show the value of the deformation potential
$\gamma$, normalized
such that $\gamma(\alpha=1) = 1$. 
The theoretical basis for this calculation can be found in Ref. \CITE{defpot}.
The deformation potential
decreases with increasing  $\alpha$. This effect can
easily be explained.
The number of nearest neighbors of the defect strongly depends on
its size and decreases with decreasing size.
This can be easily visualized for
a regular hexagonal lattice in two dimensions where particles
on interstitial sites are surrounded by three particles whereas
the coordination number of a regular lattice site
is six. Since the deformation potential
is related to the interaction of the defect with the environment
it is not surprising
that the deformation potential strongly depends on the number of
nearest neighbors.

\section{Extension of the tunneling model}
 In the  STM  it is assumed that the
asymmetry energy between two wells, $\epsilon$, and the tunneling
parameter, $\lambda = (d/\hbar)\sqrt{2mV}$, are uniformly distributed over a
wide range,
\begin{equation}\label{1}
P(\lambda,\epsilon)\, d\lambda\, d\epsilon = P_0 \, d\lambda\, d\epsilon
\end{equation}
with a constant $P_0 \approx 10^{44} -  10^{46} $~J$^{-1}$~m$^{-3}$ for most
glasses.
Here, $m$ represents
the mass of the tunneling unit, and
$d$ the distance between the minima of the double-well potential.
The tunneling parameter is related to the
tunneling amplitude, $\Delta$,  through a WKB relation
\begin{equation}\label{2}
\Delta  = \hbar \omega_0 \, e^{-\lambda},
\end{equation}
where $\hbar \omega_0$ is essentially the zero-point energy,
  and to the one-phonon relaxation rate by
\begin{equation}\label{3}
R(E,\lambda)  =  R_{\rm max}(E)\, e^{-2(\lambda - \lambda_{\rm min}(E))}
\end{equation}
where
\begin{equation}\label{3aaa}
\lambda_{\rm min}(E) = \log(\hbar\omega_0/E).
\end{equation}
 The detailed
form of the maximum rate,  $R_{\rm max}(E)$, at fixed TLS-energy, $E =
\sqrt{\Delta^2 + \epsilon^2}$,
 depends on the spectral density of vibrational
modes in the energy landscape. Usually, a deformation potential
coupling, with parameter $\gamma$, to a Debye-spectral density,
$J(\omega)\propto
\omega^3$, is imposed. The one-phonon rate then reads
$R = (\Delta/E)^2 J(E/\hbar) \coth(E/2k_BT)$ which together with (\ref{2}) and
 (\ref{3}) provides
\begin{equation}\label{3a}
R_{\rm max}(E) = c T^3 x^3 \coth x\ ,
\end{equation}
where $x = E/2k_B T$ and $c =  \gamma^2(2k_B)^3/2\pi\hbar^4\varrho
v^5$. Here,  $\varrho$ is the mass density and $v$  the averaged sound
velocity. The ensuing hyperpolic distribution for $\Delta$ and $R$,
 $P(\epsilon,\Delta) = P_0/\Delta$ or  $(\epsilon/E)P(E,R)=  P_0/2R$,
explains in particular the broad distribution of relaxation processes
in glasses on a logarithmic time scale.

The simulations suggest to extend   the tunneling model by adding a second peak
for a glass with additional defects (OH-groups) or a polymer glass
(side chain motion) [cf. Fig.~1]. In order to  restrict ourselves
to a minimum set of new parameters, we only 
characterize this second peak by its
mean value  and its width in the following model:
\begin{equation}\label{8}
P(\epsilon,\lambda) = P_0\, e^{-(\lambda-\lambda_0)^2/2\sigma_0^2}\  + \ P_1 \,
e^{-(\lambda - \lambda_1)^2/2\sigma_1^2}\ ,\qquad \lambda \ge \lambda_{\rm
min}(E),
\end{equation}
where $P_0$ and $P_1$ weights the density of the intrinsic and extrinsic TLS's.
By formulating the new distribution for the tunneling parameter, $\lambda$, we
additionally assume some relation between the potential height $V$ and
the distance between the two minima of the DWP, $d$. 
The standard assumption is $V \propto d^2$.
For $P_1 = 0$ and $\sigma_0 \gg 1$ the standard model emerges. \cite{AHV,Phil}
The parameter $\lambda_0$ determines the most probable tunneling matrix element.
For all practical applications this parameter in irrelevant and will be set
to zero hereafter. A model including only the first term
has  previously  been used by Jankowiak and Small.\cite{S0}
As we will see below, a  combination of both terms is needed in order to
explain several experiments consistently. 
These authors have also emphasized the  importance of extrinsic  TLS, which they
attribute to the presence of the chromophores in the  probe.

The ensuing distribution in $\Delta = \hbar\omega_0\, e^{-\lambda}$ then reads
after neglecting terms depending on $(1/2\sigma_0^2)
\log(\hbar\omega_0/\Delta)\ll 1$
\begin{equation}\label{9}
P(\epsilon,\Delta) = {P_0 \over \Delta} \  + \ { P_1'\,
(\hbar\omega_0)^{\nu(\Delta)}\over
\Delta^{1 + \nu(\Delta)}} \ ,
\end{equation}
where $P_1' = P_1  e^{-\lambda_1^2/2\sigma_1^2}$ and
\begin{eqnarray}\label{11a}
\nu(\Delta) = {\lambda_1\over \sigma_1^2} -
{1\over 2\sigma_1^2}\, \log (\hbar\omega_0/\Delta).
\end{eqnarray}
Identifying $\sigma_0$ with the parameter $\lambda_{\rm max}$ of the STM,
one deduces that $10 \stackrel{<}{\sim} \sigma_0 \stackrel{<}{\sim} 30$.
The value of $P_1$ depends on the concentration of the defects. For a typical
concentration
of OH-groups in network glasses, one expects $P_1 \stackrel{<}{\sim} P_0$.
In polymer glasses, however, the number of extrinsic DWP's can be larger
because of the large
number of side chains. If every side chain contributes one extrinsic DWP, and
if only
one out of $\sim 100$ monomers of the backbone  forms an intrinsic DWP
\cite{HS}, we find
as an upper bound that $P_1 \stackrel{<}{\sim} 100\,  P_0$.
More difficult is the estimation of mean value and the width of the second
peak.
For $\alpha = 1.6$  one observed numerically
an average value of $d \approx 0.6\, \sigma$. Together
with the average potential height $V_{av} = 800$~K
and using the  average atomic mass of NiP one may estimate
$\lambda_1 \sim 35$ and $\sigma_1 \sim 25$.
However, for any realistic applications $\lambda_1$ (or $V_{av}$)
 and $\sigma_1$ depend on the exact nature of the defects and hence should be
treated as adjustable parameters. Due to the observation of several absorption
maxima below the glass transition temperature,\cite{po1,po2} a more
realistic model should contain more than one additional peak in the 
 distribution of barrier heights. Since we are interested in  low temperature 
properties, it is sufficient to include  only the lowerst
addition barrier peak in our model.

\section{Comparison to experiment}
\subsection{Sound attenuation in  OH-doped network glasses and polymers}
A possible method to measure sound absorption is to clamp a  small glassy plate
on one end and to drive it electrostatically  to forced vibrations on its free
end, the so-called vibrating reed technique. \cite{Hunk,N1,Topp}  
The measurement of  the resonance frequency and the amplitude
of the plate then determines the acoustic loss and the variation of the sound
velocity.

Measurements of the elastic properties of glasses---for example the 
internal friction, $Q^{-1}$---have proven to be a powerful method for the
study of tunneling states below $\sim 5$~K, where it is assumed that 
relaxation occurs with the one-phonon
rate (\ref{3}). At higher temperatures, typically between 
10~K and room temperature, the elastic measurements provide
information about the distribution of potential barrier heights for
thermally activated reorientations,\cite{NH}
\begin{equation}\label{AR}
\tau = \tau_0 \, e^{V/k_BT}
\end{equation}
with $\tau_0 \approx  10^{-13}$~s, 
 of atoms and molecules. Assuming a simple
oscillator model for the individual wells of a DWP, we have $V\propto d^2$ and
\begin{equation}
\Delta = {2E_0\over \pi} e^{-\lambda},
\end{equation}
where $E_0$ is the zero-point energy of the oscillator, and
\begin{equation}\label{LV}
 \lambda  = {V\over E_0}.
\end{equation}
With this, the distribution (\ref{8}) can be rewritten yielding
\begin{equation}\label{DV}
P(\epsilon,V) = {P_0\over E_0}\, 
e^{-(V-V_0)^2/2\widetilde{\sigma}_0^2}\  + \ {P_1\over E_0} \,
e^{-(V - V_1)^2/2\widetilde{\sigma}_1^2}
\end{equation}
with $V_i = \lambda_i E_0$ and $\widetilde{\sigma}_i = \sigma_i E_0$,
and  the  internal friction, $Q^{-1}$, 
can be calculated with the formalism developed in Ref. \CITE{Hunk}.

Comparing the data for dry and wet $a$-B$_2$O$_3$ in Figs. 7 and 8
of Ref. \CITE{N1}, one clearly sees that the sound absorption is strongly
effected by the presence of OH-impurities. For the dry sample a broad absorption
peak is observed around $\sim 50$~K, while for the wet sample a shoulder
between 30 and 80~K is followed by a narrow  peak at $\sim 160$~K.
Similar  observations have been made in Ref. \CITE{KK}, where the dry
$a$-B$_2$O$_3$ sample provided one broad peak whereas the wet sample
showed two maxima. The authors in these references  concluded that
intrinsic DWP's  cause a  loss  peak at lower and OH-induced, 
i.e., extrinsic DWP's
a peak at higher temperatures.
Independent of any theoretical framework,  these  experiments directly
indicates that the barrier heights of these OH-related DWP's must be higher
than those of DWP's related to the glass itself.  Furthermore,
the mere fact that approximately 1\% OH-content can have
such a significant influence
on the relaxation properties shows that OH-defects have
to be very efficient in prompting the formation of DWP's.

In Fig. 6, we have plotted  the internal friction  data 
of   wet  $a$-B$_2$O$_3$ sample  together with a fit from our model
(\ref{DV}) and\cite{PHL}
\begin{equation}
Q^{-1} = {\gamma^2 \over \varrho v^2 k_B T} \int d\epsilon \int dV\ P(\epsilon,V)\ 
 {\rm sech}^2(\epsilon/2k_BT) \ {\omega\tau \over 1 + (\omega\tau)^2},
\end{equation}
where $\tau$ is given by  the one-phonon rate (\ref{3}) or the Arrhenius rate (\ref{AR})
depending on whether the temperature is in/below the plateau regime 
or above, respectively. 
 The data are  from Ref. \CITE{N1}.
The solid line shows the contribution of intrinsic and extrinsic DWP's;
the dashed line the contribution of intrinsic DWP's.
The parameter values are given in the figure caption.
 We see  that elastic measurements
are a powerful tool to determine the barrier heights distribution,
 and that our computer simulation based
model (\ref{DV}) accounts well for the influence
of OH-defects in network glasses.

Let us now discuss the elastic properties of polymers.

Nittke et al. \cite{Pobell} have recently measured the low temperature 
($< 50$~K) elastic properties  of  the polymer glasses
polymethylmethacrylate (PMMA) and polystyrene (PS) at
$\omega  = 2\pi\, \times \, 240,\ 535$  and 3200~Hz, following 
earlier measurements
by Federle and Hunklinger \cite{FH} at $\omega = 2\pi\, \times \,$15~MHz 
and Geis, Kasper, and Hunklinger\cite{po3}  
at $\omega = 2\pi\, \times \,$430~Hz  in PMMA.
In the latter work, also the high temperature absorption between 50 and 300~K 
has been measured following previous work, cf. Refs. \CITE{po1,po2,po4}.  
Below $\sim 30$~K the acoustic loss of PMMA and 
PS shows the typical temperature
dependence of network glasses like SiO$_2$: the plateau 
merges in the thermally actived relaxation 
peak around $\sim 5$~K. The missing first relaxation peak
of the acoustic loss in PMMA at  $\omega = 2\pi\, \times \,$535~Hz (cf. Fig. 12
in Ref. \CITE{Pobell} or Fig. 4 in Ref. \CITE{po3})
can well be explained by the low applied frequency $\omega/2\pi$ and a small
value for the cutoff, $V_{\rm max}$,  in the barrier distribution of 
the STM-states. \cite{cut}
However, above $\sim 30$~K both polymers behave anomalously  compared 
to (undoped)
network glasses by exhibiting a strong increase in their acoustic loss.
Early measurements show that this increase is 
  characteristic for  polymers and 
continues up to the glass transition temperature with
various ($\delta, \gamma, \beta$) relaxation peaks superimposed 
to it.\cite{po1,po2}
 Recently, multidimensional NMR measurements have proven 
that the molecular origin of the $\beta$ relaxation in PMMA
at 330~K (10~Hz)   is a  large-amplitude flip of the methacrylate side 
group around the C--C bond and a concomitant  main chain torsion.\cite{Rohr} 
 From the evidence at present available the nature of 
the groups responsible for the low  temperature $\delta, \gamma$ 
peaks which have been observed in different polymers 
 is less clear, however, they have traditionally been attributed to 
thermally activated  rotation of sub groups within the side chains. 
Examples are  the 
$n$-alkyloxycarbonyl side group rotations  of 
the  polymethacrylates, see Figs. 3 and
4 in Ref. \CITE{po2}. In PMMA these side groups do not exist  and
absorption peaks observed at temperatures 
below the $\beta$ peak \cite{po3} are thought  to be due to the presence 
of small amounts of dibutylphtalate (DBP)---a plasticizer---or higher
polymethylacrylate esters.  Studies on purified samples,
which revealed a pure exponetial increase, $Q^{-1} \propto e^{aT}$,
between $\sim 30$ and $\sim 300$~K, support this interpretation.\cite{po1,po2}
However, even with a broad distribution, it is not clear whether   
 the  exponential increase is connected 
with the $\beta$ process. It is conceivable that 
further molecular dynamics occurs in the  intermediate temperature regime.

In any case we learn from these experiments that (1)
the  similarity  between   sound absorption  in OH-doped network
glasses and in polymer glasses may justify to think of side chain
as  ``defects'', because  both hydroxcyl groups and side groups give
rise to well-defined localized motion within  DWP's with high 
activation barriers, and thus constitute additional
degrees of freedom; (2) the energy landscape of polymers is structured at
higher barrier heights: in addition to the uniformly distributed 
low barriers being responsible for the plateau region below $\sim 6$~K,
several tiers of high barriers with typical activation energies can exist
depending on the chemical compound of the polymer.

\subsection{Hole burning in polymer glasses and proteins  on ultralong time
scales}
 In a  hole burning experiment  the probe
is doped with a large number of guest molecules having resonance
frequencies in the optical range. Due to the disordered structure of the host
each chromophore experiences a slightly different environment which detunes
its resonance frequency; the absorption line becomes inhomogeneously
broadened, accordingly. A fraction of these chromophores is
photochemically or photophysically removed from the absorption line by
laser irradiation. The inhomogeneous line then
shows a narrow hole at the laser irradiation frequency. Each chromophore is
interacting with local rearrangements of atoms in the host which
conveniently can be described in the TLS picture. As a result, the
resonance frequency of the guest molecule  performs spectral jumps and its
linewidth becomes broadenend. This process, which is called ``spectral
diffusion'',  results in a partial refillment and, thereby, in a broadening
of the hole. Hence, the time dependence of the hole width is a mirror of the
relaxation processes in the host, and, accordingly, a tool to investigate
the local dynamics of the host. In proteins and glasses the results
can be interpreted by the concept of an energy landscape.

In recent years this technique has been used extensively by various groups
on glasses and  proteins. \cite{FS,Haar1,Haar2,Haar3,FR3,FR4,FR1,FR2,BF,Vol1,Vol2}
Most measurements on glasses for times $t \le 3$~h
confirm the STM,
which predicts according to $(\epsilon/E) P(E,R) = P_0/2R$  and
\begin{eqnarray}\label{4}
\Delta\Gamma(t) &=& {\pi^2\over 3\hbar} \langle C\rangle
\int dE\,   {\rm sech}^2{E\over 2k_B T} \int dR\,
 {\epsilon\over E}\, P(E,R) \left(e^{-R t_0}
- e^{-R t}\right)
\end{eqnarray}
a logarithmic growth
\begin{equation}\label{5}
\Delta\Gamma(t)  \equiv \Gamma(t) -
\Gamma(t_0) = {\pi^2\over 3\hbar} P_0 \langle C\rangle k_B T \,
\log(t/t_0)
\end{equation}
of the hole width  after some reference time $t_0$.  \cite{BF,HW,BH,Re,SS}
Here, $\langle C\rangle$ is the chromophore-TLS coupling strength. This
accords with the broad distribution of relaxation times on logarithmic time
scales.

Recently, Haarer and coworker  \cite{Haar1,Haar2,Haar3} have 
performed hole burning on purified PMMA at 0.5, 1 and 2~K up to 
extremely long times (from 10~s to 10~days),
see data points in Fig.~7.
They found a $\log(t)$-dependence up to ca. 3~h,
but  a $\sqrt{t}$-behavior between 3~h and 10~days.
Furthermore,  Friedrich and coworker  \cite{FR3,FR4} have performed  equivalent
measurements on proteins at 100~mK and 4~K.  In Ref. \CITE{FR3}, these
 authors did    photochemical hole burning on a
glycerol/water glass  over 44~h with and without  the protein
myoglobin doped into the sample, see data points in Fig.~8.
 Recently, they also studied the protein cytochrome c
in a glycerol/dimethylformamide glass\cite{FR4} for 300~h, 
see data points in Fig.~9.
 The result was: (i) spectral diffusion broadening is always less in the
protein than in the glass; (ii) the glass showed a logarithmic
time dependence within the first 3~h, while there was practically no
relaxation in the protein over this period; (iii) after about 3~h there is
a strong nonlogarithmic increase in the line broadening for both the glass
and the protein; (iv) contrary to the glass, the protein showed no aging effect
indicating a gap in the relaxation rate between 1~s and $\sim 3$~h.
 The protein and the glass data
could be fitted with an empirical {\it ad hoc} ansatz for
the distribution function
\begin{equation}\label{6}
P(\epsilon,\Delta) = P_0 \left[ {1\over \Delta} + {A\over \Delta^2}\right]\  ,
\qquad  A = {\rm constant},
\end{equation}
resulting in a growth composed of a superposition of a $\log(t)$- and a
$\sqrt{t}$-term.
For the 0.5  and 1~K  data in PMMA a   single consistent parameter set
with $A/k_B = 10^{-7}$~K has been used. \cite{Haar1,Haar2}
For the 2~K data in PMMA,
however, a 60\% smaller value for the weight parameter
 $A$ was needed.\cite{Haar3}
For the protein data the question rises whether 
there is a $\log(t)$-term or not as
the data can already be reproduced by the $\sqrt{t}$-term. This observation
motivated the authors in Ref. \CITE{FR4} to discuss a possible
alternative interpretation of  the nonlogarithmic dynamics which is related to
classical diffusion in the energy landscape.

In both experiments the distribution function
(\ref{6}) has been motivated by recent publications focusing
on the interaction of TLS's in glasses,
\cite{YL,Copp,BK} in particular by the scenario
 of coherently coupled pairs of TLS's, which was
invented by Burin and Kagan. \cite{BK}
In this formulation the second term in (\ref{6}) stems from
pairs of single TLS's which are coupled by a
(resonant) up-down transition. \cite{BK,N2} If the single TLS's are distributed
as in the STM [first term in Eq. (\ref{6})], a $1/\Delta^2$-distribution
emerges for the pairs. 
In Ref. \CITE{N2}, a quantitative test of this picture has been performed
with the conclusion that this strongly modified TLS model could,
 in principle, explain the deviation from the $\log(t)$-dynamics.
At higher temperatures  pairs are expected
to break up due to thermal fluctuations which destroy the coherence of the
coupling. It has been suggested that this is the physical reason for the decrease
 of the weight  factor $A$ seen in the PMMA data at 2~K. \cite{Haar3}

In the following, we will propose  an alternative  explanation, 
which is motivated by the acoustic measurement on polymers, and
has no restriction towards  higher temperatures.

Instead of considering TLS-flip-flop processes as the physical cause
of spectral diffusion, we take barrier crossings events. On the time scale
of seconds and larger, both picture are of course equivalent 
due to the big ratio $\epsilon/\Delta$ of the relevant TLS or DWP's.
We assume that at 1~K the barrier crossing process occurs via tunneling
with the one-phonon rate, Eqs. (\ref{3})--(\ref{3a}), and 
consider the tunneling parameter $\lambda$, which 
in the simple oscillator model (\ref{LV}) is equivalent
 to the barrier height $V$. 
Because of  the observation that the broadening
 significantly  increases at 1~K between $t_c = 10^3$ and  $10^4$~s, and that
the distribution $P(\Delta)\propto 1/\Delta^2$ well accounts for this
phenomenon, we learn that  the distribution $P(\lambda)$ increase
exponentially around some value $\lambda_c \gg 1$.
We may estimate this value from the relation $R(1\, {\rm K}) t_c \sim 1$ which
provides with $R_{\rm max}(1\, {\rm K})
 \approx 10^{10}$~s$^{-1}$ that  $\lambda_c -\lambda_{\rm min}(1\, {\rm
K})\approx 15$.
A natural way to interprete
this increase in $\lambda$ is to assume that {\it the energy landscape of a
polymer glass is not structureless, but that it comprises
high barriers in addition to structurelessly distributed
 lower barriers within each basin}---as observed in the acoustic experiments
on polymer glasses. This is the picture which Frauenfelder
suggested for the energy landscape of proteins. \cite{F1,F2,F3}
We may identify transitions
between deep basins with degrees of freedom (rotations or switching) located
at the side chains of the polymer glass or protein. 
The STM-like states may be associated
with the more collective dynamics along the backbone or the network.
 One might think that already the backbone dynamics
  of polymers gives rise to dynamics with high barrier heights, since
  this dynamics is related to conformational transitions.
  However, it has been shown for some polymers that conformational
   transitions occur on the same time scale as the $\alpha$ relaxation and
  thus can be neglected at temperatures far below $T_g$. \cite{Zemke} The
reason is  that conformational transitions involve highly
 cooperative dynamics in
   contrast to more localized processes we are here interested in and
   which are still active below $T_g$. 

The difference between  protein and  glass then simply lies in the amount of
disorder
and organization in the energy landscape.
Though deep basins separated by saddle points
exist in both materials, in proteins the distribution of small (STM-like)
barriers in the basins is restricted to smaller values, i.e., $\sigma_0({\rm
protein})\ll
\sigma_0({\rm glass})$, or, more likely, itself composed of
many well-separated distributions yielding the hierarchical picture of
Frauenfelder
with STM-like states on the lowest tier. Indeed,
recent stimulated echo experiments by Thorn Leeson and Wiersma can be
interpreted
in this way. \cite{TW} Hence, whereas many different tiers
can exist in proteins, the glass is less organized:
the structurelessly distributed STM-states extend to much higher barriers,
  and only the passing from one basin
to the next gives rise to an (in the sense of the STM) anomalous behavoir.
Hence, we suggest a connection between the low temperature 
($\stackrel{<}{\sim} 100$~K)
relaxation peaks observed in elastic measurements with the
nonlogarithmic line  broadening observed in spectral hole burning
after $10^3 - 10^4$~s in the temperature regime around 1~K. 
A  crossing of barriers as high as $\stackrel{<}{\sim} 1000$~K
by {\it tunneling} at 1~K is only possible because of the extreme
long waiting time of $10^3 - 10^6$~s.
At higher temperatures ($\stackrel{>}{\sim} 5$~K)
the one-phonon rate  should gradually be 
replaced by the Arrhenius rate (\ref{AR}), which will change
the temperature dependence, but not the time dependence of the hole broadening.

The simplest way to quantify our ideas   is to calculate the hole broadening
from  Eq. (\ref{8}) and compare with the experimental data of Refs.
\CITE{Haar1,Haar2,Haar3} and
\CITE{FR3,FR4}. In the Appendix we give details of the derivation. We find for
typical values
of the shortest time a hole can be read, $t_0 \sim 1$~s,
\begin{equation}\label{12}
\Delta \Gamma (t) = {\pi^2\over 3\hbar}  \langle C\rangle\, k_B T
\Bigm[ P_0 \sigma_0 \sqrt{2\pi} \,  f^{(1)}(t) \ +\   P_1\sigma_1 \sqrt{2\pi}\,
f^{(2)}(t)\Bigm],
\end{equation}
where
\begin{eqnarray}\label{ft1}
f^{(1)} (t) &=&  {\rm erf}
\left(\frac{(1/2)\log(KTt)}{\sigma_0\sqrt{2}}\right)\ - \
 {\rm erf}\left(\frac{(1/2)\log(KTt_0)}{\sigma_0\sqrt{2}}\right), \\[0.2cm]
\label{ft2}
f^{(2)} (t) &=&  {\rm erfc}\left({\lambda_1 -
(1/2)\log(KT  t) \over \sigma_1 \sqrt{2}}\right) \ - \
{\rm erfc}\left({\lambda_1 - (1/2)\log(KT t_0) \over
\sigma_1\sqrt{2}} \right).
\end{eqnarray}
Here, erf$(x)$ (erfc$(x))$ is the (complementary) error function,
and $K$ has been defined in Eq. (\ref{K}).
The first term  yields  with ${\rm erf}(z) \stackrel{z\ll 1}{\longrightarrow}
(2/\sqrt{\pi}) z$   in leading order in  $(1/\sigma_0)$  the standard 
result (\ref{5}). If the barrier crossing is thermally activated with
rate $\tau^{-1} = \tau^{-1}_0 e^{-V/k_BT}$, one easily finds that
within the simple oscillator model (\ref{LV}) and (\ref{DV})
\begin{eqnarray}\label{ft3}
f^{(1)} (t) &=&  {\rm erf}
\left(\frac{k_B T\log(t/\tau_0)}{\widetilde{\sigma}_0\sqrt{2}}\right)\ - \
 {\rm erf}\left(\frac{k_B T\log(t_0/\tau_0)}{\widetilde{\sigma}_0\sqrt{2}}
\right), \\[0.2cm]
\label{ft4}
f^{(2)} (t) &=&  {\rm erfc}\left({V_1 -
k_BT\log(t/\tau_0) \over \widetilde{\sigma}_1 \sqrt{2}}\right) \ - \
{\rm erfc}\left({V_1 - k_BT\log(t_0/\tau_0) \over
\widetilde{\sigma}_1\sqrt{2}} \right),
\end{eqnarray}
with $\widetilde{\sigma}_i = E_0 \sigma_i$, 
which, as mentioned above, changes the temperature but not the time dependence
of the line width.

 In Figs.~7 to 9,  we compare
Eq. (\ref{12}) - (\ref{ft2})  with the experimental data for  PMMA,
glycerol/water glass and myoglobin, and cytochrome,  respectively.
The parameter values are given in Table I and II. 
In the glasses PMMA and glycerol/water glass,
 logarithmic line broadening  occurs below 
$\sim 10^4$~s, as predicted by  the STM.  
 The experimentally detected 
algebraic $\sqrt{t}$-time dependence applies approximately
 between $10^4$ and $10^6$~s.
As is clearly demonstrated in Fig.~7, it 
arises from a superposition of the two terms in Eq. (\ref{12}).
Hence, the experimentally found exponent $1/2$ has no physical significance
in this model. The protein data can be fitted
without a $\log(t)$-term (i.e., $P_0 = 0$) as the dashed lines in Fig.~8 and 9
illustrate. In any case the contribution of the $\log(t)$-term is very small
as can be inferred from the dash-dotted curve in Fig.~9.
This is consistent with the absence of aging observed in Ref. \CITE{FR4}
(see data points corresponding to different equilibration times in Fig.~9),
which indicates a gap in the distribution of relaxation times between
$\sim 1$~s and 3~h.  The hole broadening
of  glycerol/water glass and myoglobin embedded in the glycerol/water glass
could not be fitted with the same set of parameters, which clearly
indicates the protein comprises itself TLS's which dephase the chromophore.
The numerical values for $K$, $P_0 \langle C\rangle$ and $\sigma_0$
are in the range known  from other experiments.
  The numerical values for the new parameters
$\lambda_1$, $\sigma_1$, and $P_1/P_0$ lie  well within the
range we estimated from the numerical simulations.
Using the parameter set for PMMA in  Table I for the acoustic
measurement of Refs. \CITE{po3,Pobell} is difficult because we do not
know the exact relation between $V$ and $\lambda$. However, assuming
$E_0({\rm PMMA}) \sim 10 - 20$~K we see that the increase 
in the sound absorption in PMMA should be at a lower temperature 
than for the OH-doped Boroxid glass---as  observed in the experiment.

\subsection{Comparison of elastic and optical properties of PMMA
and experimental test} 
Our picture of a structured barrier heights distribution
seems to be consistent  with elastic and optical measurements  
on polymers. However,  in PMMA the puzzle remains that purified samples 
contain no sound absorption peak
in the temperature regime below the $\beta$ peak ($\sim 330$~K).
 Hence, no peaks in the barrier 
heights  distribution seem to exist in  the regime below say 1000~K, 
which is the relevant order of magnitude 
to obtain a nonlogarithmic line broadening between
$10^4$ and $10^6$~s. However, one must exercise caution since the
experimentally observed exponential increase in the acoustic loss
between 20 and 300~K can hide different dynamic  molecular processes. 
Furthermore, in principle, our simulations show that
{\it any}  impurity can generate such an additional 
peak in the barrier distribution, and that defects are very effective
in prompting external DWP's. A very low concentration of defects
is sufficient to have dramatic effects. 
Hence, it is possible that the presence of impurities 
such as exchange gas atoms,
monomers, fragments, hydrocyl groups, etc., is reponsible for this process.
Also, it is unclear what the role of the chromophores may be,
 although the observation of a logarithmic line broadening in proteins,
where the chromophore only replaces a group of a {\it similar} structure (i.e.,
does not change the local configuration  of the protein), makes the presence 
of the chromophore a doubtful cause of the nonlogarithmic line broadening.
Furthermore, our simulation did not show any additional peaks in the
barrier heights  distribution function upon adding {\it large}
 impurities to the LJ-glass.

There are several possibilities for an experimental test whether 
the physical cause  of the nonlogarithmic line broadening is due to
the explanation  given in this work or due to the
TLS-TLS interaction as pointed out in Ref. \CITE{N2}. 
First, it would be interesting to know whether this effect
is universal, i.e., whether  hole burning on different polymers 
give identical results: same power law and same crossover time
from the logarithmic to the nonlogarithmic (or algebraic) behavior.
If this were true, it would hint towards TLS-TLS interaction as the
physical cause. Even in this case, however, it would be difficult to
argue in favor for this explanation at temperatures
above $\sim 5$~K. It seems so that
independently of a possible coherent TLS coupling around  1~K, the explanation
given in this work is more intuitive at higher temperatures.
Especially,  the observation of a transition from (\ref{ft1}) and 
(\ref{ft2}) to (\ref{ft3}) and (\ref{ft4}) would favor this interpretation.
 Next,  elastic measurement up to higher temperatures 
on exactly the same sample which has been used for the hole burning 
experiment, i.e., including the chromophore, would also give additional 
information. Finally, the question
how  polymer glasses without side chains, like   polyethylene (PE) 
or polybutadiene (PB), behave in hole burning at ultralong
time scales seems to be interesting. 
 Following the interpretation of the present work,
one would expect  no or a later deviation from a logarithmic
hole broadening (if $\sigma_0$ is sufficiently large)
  as compared to polymers like PMMA because of   the lack of ``defects.''

\section{Summary and Conclusions}
We have analyzed the energy landscape of
glasses containing defects. Specifically, we refer
to OH-defects. However, the experimental results
on polymers allow to speculate that also some side groups
in polymers like the carboxyl group in PMMA
may be viewed as a (generalized) defect.

The analysis of LJ glasses has revealed that already
small differences between defect and glass dramatically
change the energy landscape around this defect, giving
rise to additional extrinsic DWP's.
The main characterstics of these extrinsic DWP's are
(i)  high barrier heights, (ii) high probability
that a defect is indeed connected with an extrinsic DWP,
(iii) highly localized dynamics around this defect. As already discussed
in the Introduction (i) and (ii) are in direct agreement
with experimental observations on network-formers.
Although the results are derived for a specific model glass
we believe that this behaviour may be generic for other
types of defect/glass-systems.

This observation allows a straightforward extension of the STM which
successfully accounts for the nonlogarithmic dynamics of
hole burning experiments on polymer glasses and proteins.
In terms of the energy landscape 
these low temperature  experiments mainly probe their fine
structure. Comparing the hole burning experiments for polymer glasses
and proteins for short times leads to the conclusion that the
fine structure is more distinct for polymer glasses. However,
the presence of some high energy saddles, corresponding to localized
motion, seems to be present in both systems.
 
\section*{Acknowledgements} 
P. N.  gratefully acknowledges financial support by 
the Alexander von Humboldt  foundation.
 We would like to thank R. J. Silbey, J. Friedrich, 
D. R. Reichman, K. Fritsch, and R. Wunderlich
for discussion and providing experimental data prior to publication.
 In particular, we thank 
P. O. Pohl and Ch. Enss for valuable comments  and 
bringing references about elastic properties of
polymers to our knowledge.

\setcounter{equation}{0}
\appendix

\renewcommand{\theequation}{\Alph{section}.\arabic{equation}}
\section{}
In this Appendix we derive Eq. (\ref{12}). Starting out from Eq. (\ref{4}),
we first note that
\begin{eqnarray}\label{A1}
[\epsilon(E,R)/E] P(E,R)\, dE dR &=& [\epsilon(E,\lambda)/E] P(E,\lambda) \, dE
d\lambda \nonumber\\
&=& \left(P_0 e^{-(\lambda-\lambda_0)^2/2\sigma_0^2}+ P_1 e^{-(\lambda -
\lambda_1)^2/2\sigma_1^2}\right) \,dE d\lambda.
\end{eqnarray}
Rewriting Eq. (\ref{3}) as
\begin{equation}
t R(E,\lambda) = e^{-2[\lambda - \lambda_{\rm min}(E) - (1/2)\log(t R_{\rm
max}(E))]},
\end{equation}
we may use the step function approximation for the term $(e^{-t_0R(E,\lambda)}
- e^{-tR(E,\lambda)})$
in Eq. (\ref{4}),
\begin{equation}
\int_{\lambda_{\rm min}(E)} d\lambda \left(e^{-t_0R(E,\lambda)} -
e^{-tR(E,\lambda)}\right) ... \ \approx \
\int_{\xi(E)}^{\lambda_{\rm min}(E) +  (1/2)\log(t R_{\rm max}(E))} d\lambda \
...\ ,
\end{equation}
where
\begin{equation}
\xi(E) = {\rm max}[\lambda_{\rm min}(E),\lambda_{\rm min}(E) +  (1/2)\log(t_0
R_{\rm max}(E))].
\end{equation}
For a hole burning experiment,  we have $t_0 > 1$~s such  that at $T\sim
O(1)$~K
\begin{equation}\label{HBD}
\xi(T)  = \lambda_{\rm min}(T) +  (1/2)\log(t_0 R_{\rm max}(T)) =
 (1/2)\log(t_0 \widehat{R}(T)),
\end{equation}
where
\begin{equation}\label{K}
\widehat{R}(E) \equiv R_{\rm max}(E) e^{2\lambda_{\rm min}(E)} =   KT\, x\,
\coth x,
\end{equation}
$K = c (\hbar\omega_0/2k_B)^2$ [cf. Eq.~(\ref{3a})].
Note that, due to Eqs. (\ref{3aaa}) and (\ref{3a}), $\widehat{R}$
depends linearly on temperature.
From the first term in Eq. (\ref{A1}),  we then find with \cite{AS}
\begin{equation}\label{err}
{1\over\sigma\sqrt{2\pi}} \int_{-\infty}^x  e^{-(y - m)^2/2\sigma^2} dy  =
{1\over 2} \left( 1 + {\rm erf}
\left(\frac{x-m}{\sigma\sqrt{2}}\right)\right),
\end{equation}
where ${\rm erf}(z)$ is the error function,  
that (after setting $\lambda_0 \equiv 0$)
\begin{eqnarray}\label{Ax}
\Delta\Gamma^{(1)}(t)  &=& (\pi^2/3\hbar) \langle C\rangle P_0 \int_0^\infty
dE\ {\rm sech}^2(E/2k_BT)\ \times \nonumber\\
&\times&  {\sigma_0 \sqrt{2\pi}\over 2}
\left\{{\rm
erf}\left(\frac{(1/2)\log(t\widehat{R}(E))}{\sigma_0\sqrt{2}}\right)\ - \
 {\rm erf}\left(\frac{\xi(E)}{\sigma_0\sqrt{2}}\right)
\right\}.
\end{eqnarray}
From the second  term in Eq. (\ref{A1}), we find
with (\ref{err}) and the definition of the  complementary  error function
${\rm erfc}(z) = 1 + {\rm erf}(-z)$ that
\begin{eqnarray}\label{Axx}
\Delta\Gamma^{(2)}(t)  &=& (\pi^2/3\hbar) \langle C\rangle P_1 \int_0^\infty
dE\ {\rm sech}^2(E/2k_BT)\ \times \nonumber\\
&\times&  {\sigma_1 \sqrt{2\pi}\over 2}
\left\{{\rm erfc}\left(\frac{\lambda_1 -
(1/2)\log(t\widehat{R}(E))}{\sigma_1\sqrt{2}}\right)
\ - \ {\rm erfc}\left(\frac{\lambda_1 -
\xi(E)}{\sigma_1\sqrt{2}}\right)\right\}.
\end{eqnarray}
Upon noting that  $\widehat{R}(E)$ is approximately  
independent of $E$ for $E < 2k_B T$, we find
   Eq. (\ref{12}) in the hole burning  time regime (\ref{HBD}).

\subsection*{Figure captions}

\noindent FIG. 1.
The probability to find extrinsic and intrinsic DWP's by
our search algorithm. As discussed in the text the
probability for intrinsic DWP's has been scaled such
that for $\alpha = 1$ both probabilities agree. The observation
that no extrinsic DWP's have been found for $\alpha = 0.7$
is indicated by the arrow. Note the logarithmic scale for
the probability.\\

\noindent FIG. 2.
The average effective mass $p$ in dependence of $\alpha$.\\

\noindent FIG. 3.
The distribution of potential heights for $\alpha = 1.0$ and $\alpha = 1.6$.\\

\noindent FIG. 4.
The average potential height $V$ in dependence of $\alpha$.\\

\noindent FIG. 5.
The average deformation potential $\gamma$ in dependence of $\alpha$.
The data have been normalized such that $\gamma(\alpha = 1) = 1$.\\

\noindent FIG. 6. 
Internal friction, $Q^{-1}$, of the OH-doped network glass
$a$-B$_2$O$_3$. The data are from Ref. \CITE{N1}.
The solid line is a fit with the distribution (\ref{8}); 
the dashed line  line shows 
the contribution of the internal DWP's.
 The parameters are $C \equiv P_0 \gamma^2/\varrho v^2 
= 3.6 \times 10^{-4}$, $E_0/k_B = 18$~K,
$\widetilde{\sigma}_0 = 450$~K, $P_0/P_1 = 0.22$, $V_1 = 2000$~K,
and $\widetilde{\sigma}_1 = 900$~K. \\

\noindent FIG. 7.  Hole broadening in the polymer glass  PMMA at
2~K (upper curve), 1~K (middle curve), and 0.5~K
(lower curve). The experimental data are from Ref. \CITE{Haar2,Haar3}.
The solid lines are fits with Eq. (\ref{12}) with parameters given in Table I;
the dashed line depicts the contribution of the first term, the dash-dotted
line the contribution of the second term in Eq.  (\ref{12}).
A nearly algebraic line broadening between $10^2 - 10^4$~min emerges from the
superposition of both terms.\\

\noindent FIG. 8.  Hole broadening in glycerol/water glass
(upper curve) and the protein myoglobin (lower curve).
The experimental data are from Ref. \CITE{FR3}.
The solid and dashed line are fits
with Eq. (\ref{12}) with parameters given in Table I and II.
The protein data have been fitted without a $\log(t)$-term ($P_0 = 0$).\\

\noindent FIG. 9.  Hole broadening in the protein cytochrome c.
The experimental data are from Ref. \CITE{FR4};  time between cooling and
burning:
94~min $(+)$, 4398~min $(\circ)$, and 10093~min $(*)$.
The full and dashed line are fits
with Eq. (\ref{12}) with both terms (full line) and only the second term
(dashed line).
Parameters given in Table I and II.
The dash-dotted line depicts  a possible contribution
of a $\log(t)$-term.


\subsection*{Tables}

$$
\begin{tabular}{l|cccccc}
\multicolumn{7}{l}{Table I: Parameters for PMMA, glycerol/water glass and
cytochrome}\\
\multicolumn{7}{l}{$\qquad\quad\ $ (intrinsic + extrinsic DWP's):}\\
 \hline\hline
   & $\lambda_1$ & $\sigma_0$ &$\sigma_1$& $P_1/P_0$ &$P_0\langle
C\rangle$&$K$ [K$^{-1}$s$^{-1}$]  \\ \hline\hline
PMMA          &   20   &  30   & $\sqrt{1.3}$ & 8 &$0.95\times 10^{-5}$&$5
\times 10^{11}$  \\\hline
glycerol/water  &  20    & 20   & $\sqrt{1.7}$& 28 &$3.49 \times 10^{-5}$&
$3.55 \times 10^{11}$ \\\hline
cytochrome     &19.7    &20    & 1          &  50 & $0.35  \times 10^{-5}$&
$4.1  \times 10^{10}$\\\hline\hline
\end{tabular}
$$
\vspace{1cm}
$$
\begin{tabular}{l|ccccc}
\multicolumn{6}{l}{Table II: Parameters for myoglobin and cytochrome}\\
\multicolumn{6}{l}{$\qquad\quad\ \ $ (extrinsic DWP's):}\\
 \hline\hline
 &  $\lambda_1$ &$\sigma_1$&$P_0$ &$P_1\langle C\rangle$&$K$
[K$^{-1}$s$^{-1}$]  \\ \hline\hline
 myoglobin  & 20  & $\sqrt{3}$ & 0 &$62.4 \times 10^{-5}$&$6.6\times 10^{10}$
 \\\hline
cytochrome  & 19.7 & 1         & 0 &$17.3 \times 10^{-5}$&$4.1\times 10^{10}$
 \\ \hline\hline
\end{tabular}
$$

\newpage
\subsection*{Figures}

\noindent FIG. 1: Heuer et al.

\begin{center}
              \epsfig{file=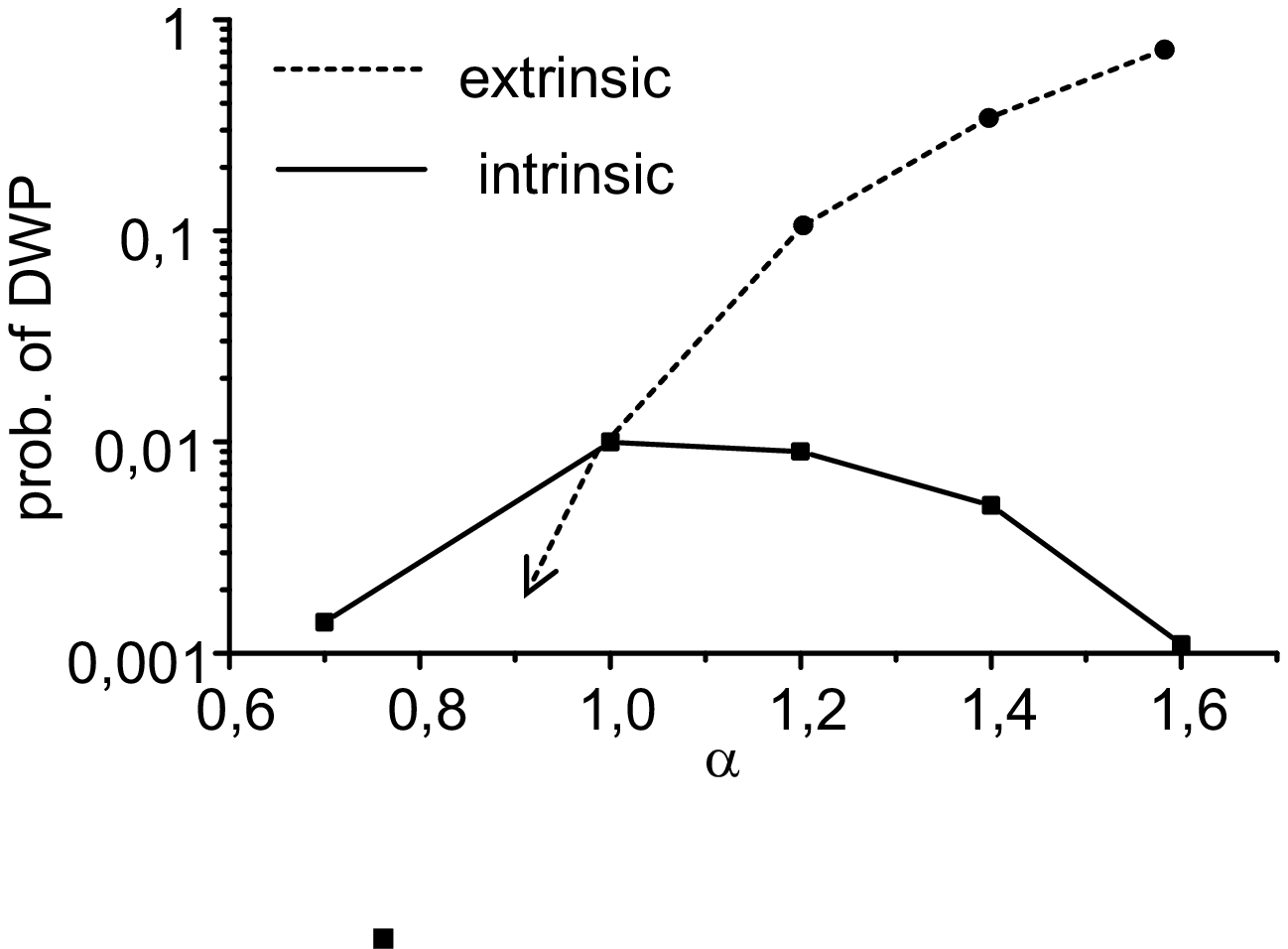}
\end{center}

\noindent FIG. 2: Heuer et al.
\begin{center}
              \epsfig{file=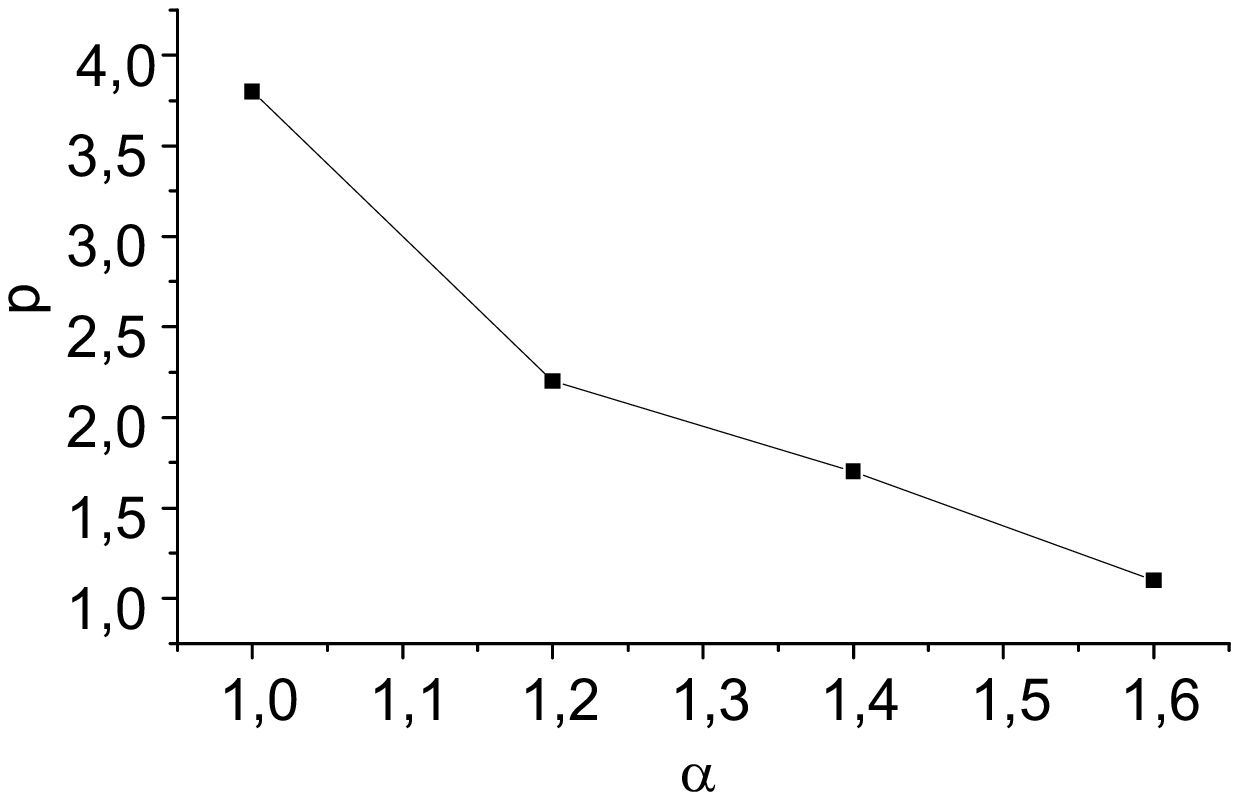}
\end{center}

\newpage

\noindent FIG. 3: Heuer et al.

\begin{center}
              \epsfig{file=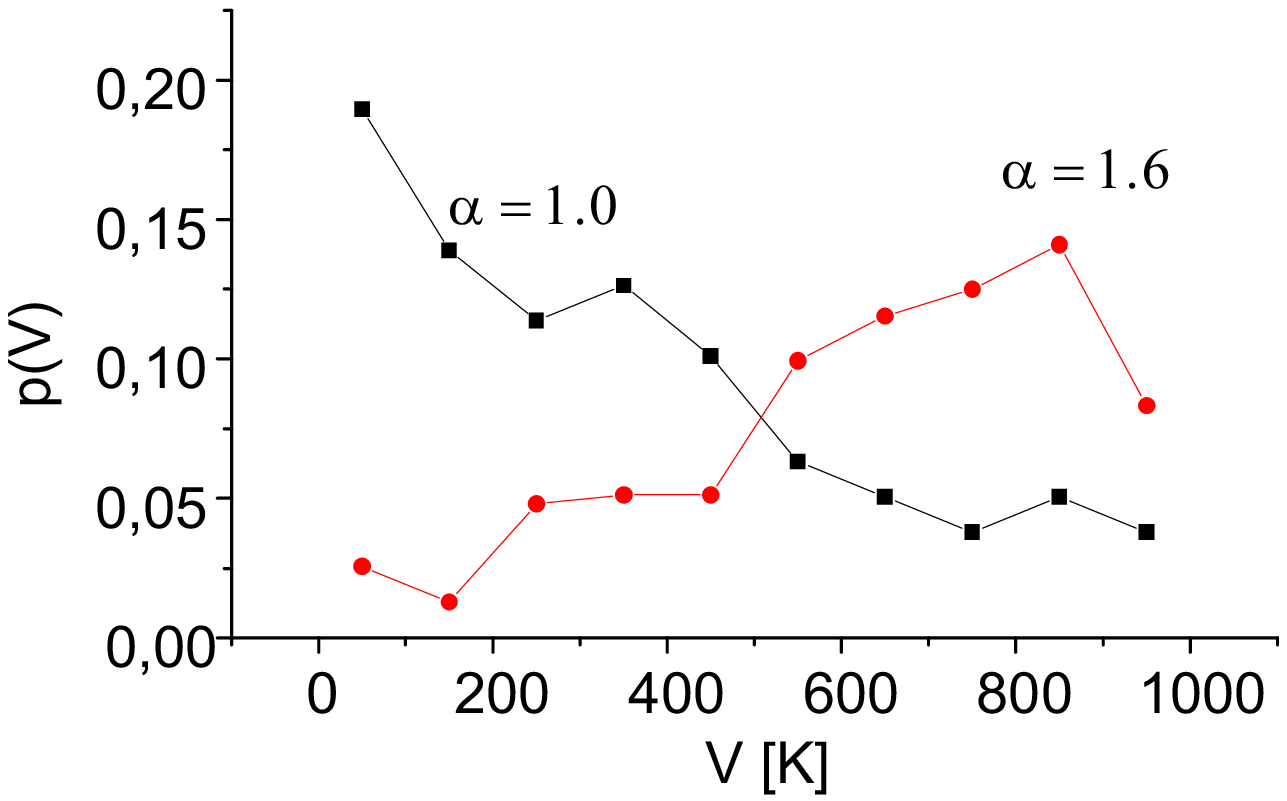}
\end{center}

\noindent FIG. 4: Heuer et al.

\begin{center}
              \epsfig{file=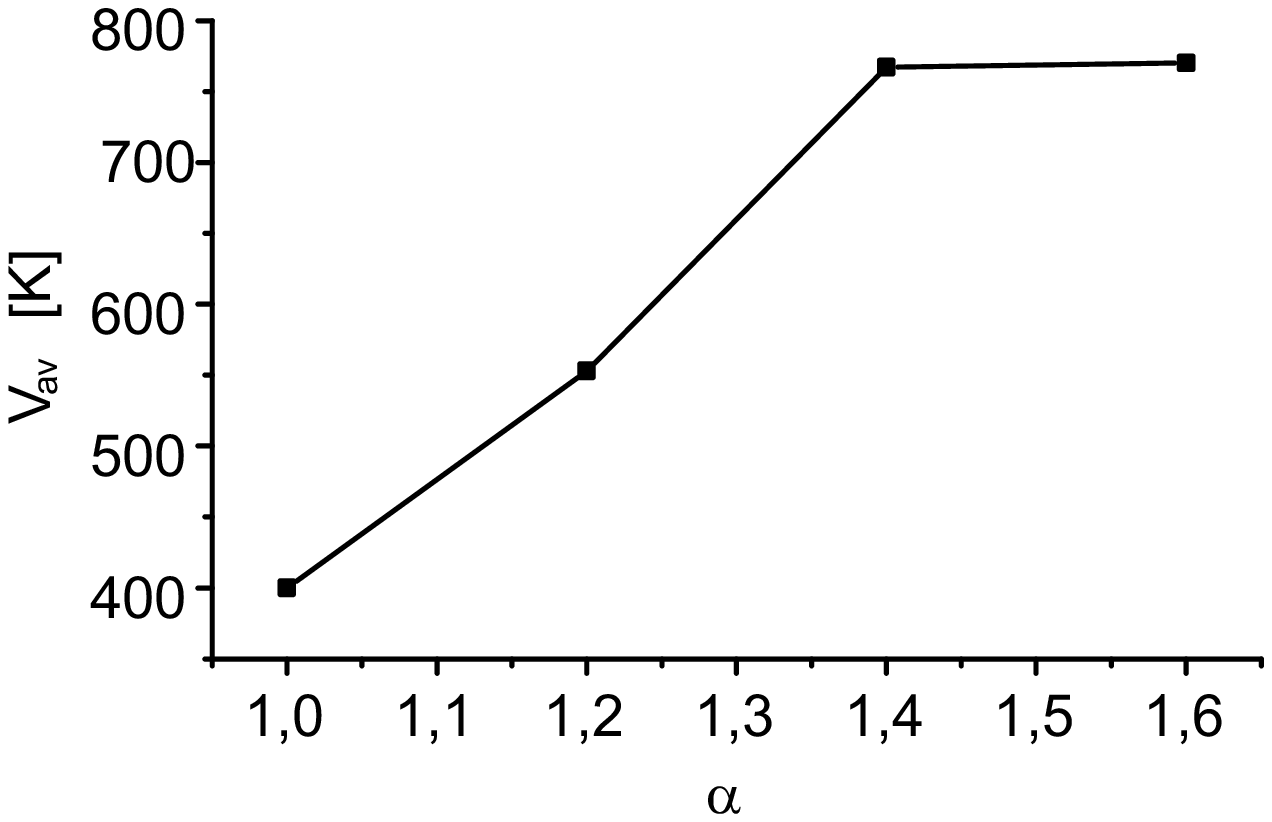}
\end{center}

\newpage

\noindent FIG. 5: Heuer et al.

\begin{center}
              \epsfig{file=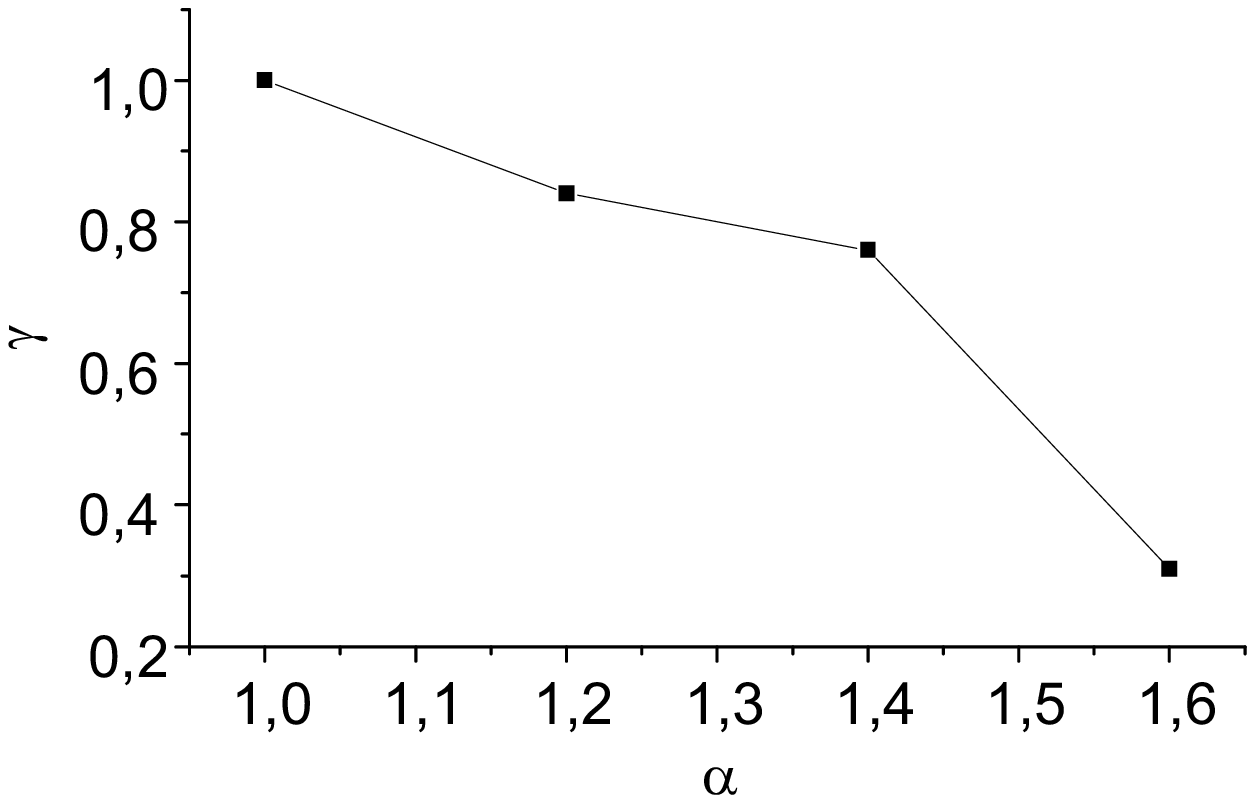}
\end{center}

\noindent FIG. 6: Heuer et al.

\begin{center}
              \epsfig{file=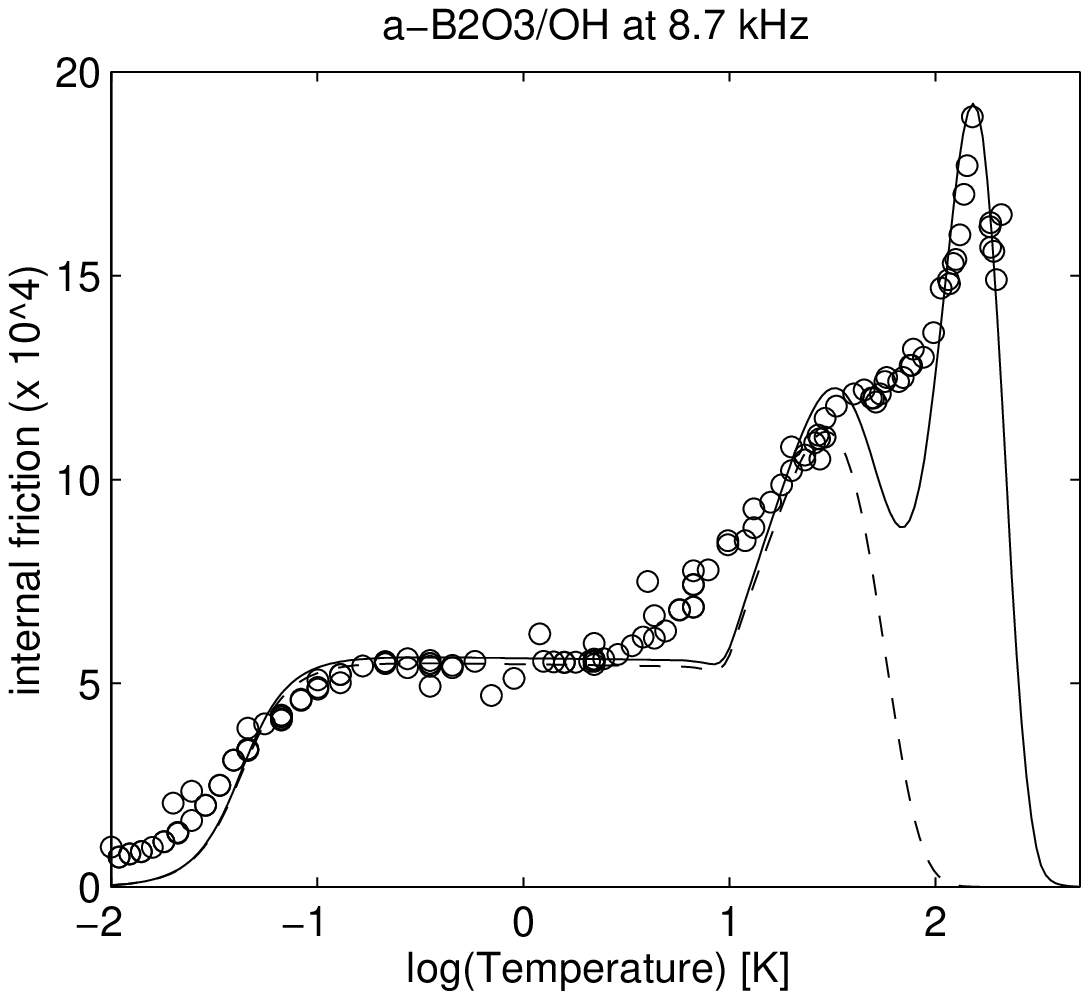}
\end{center}

\newpage

\noindent FIG. 7: Heuer et al.

\begin{center}
               \epsfig{file=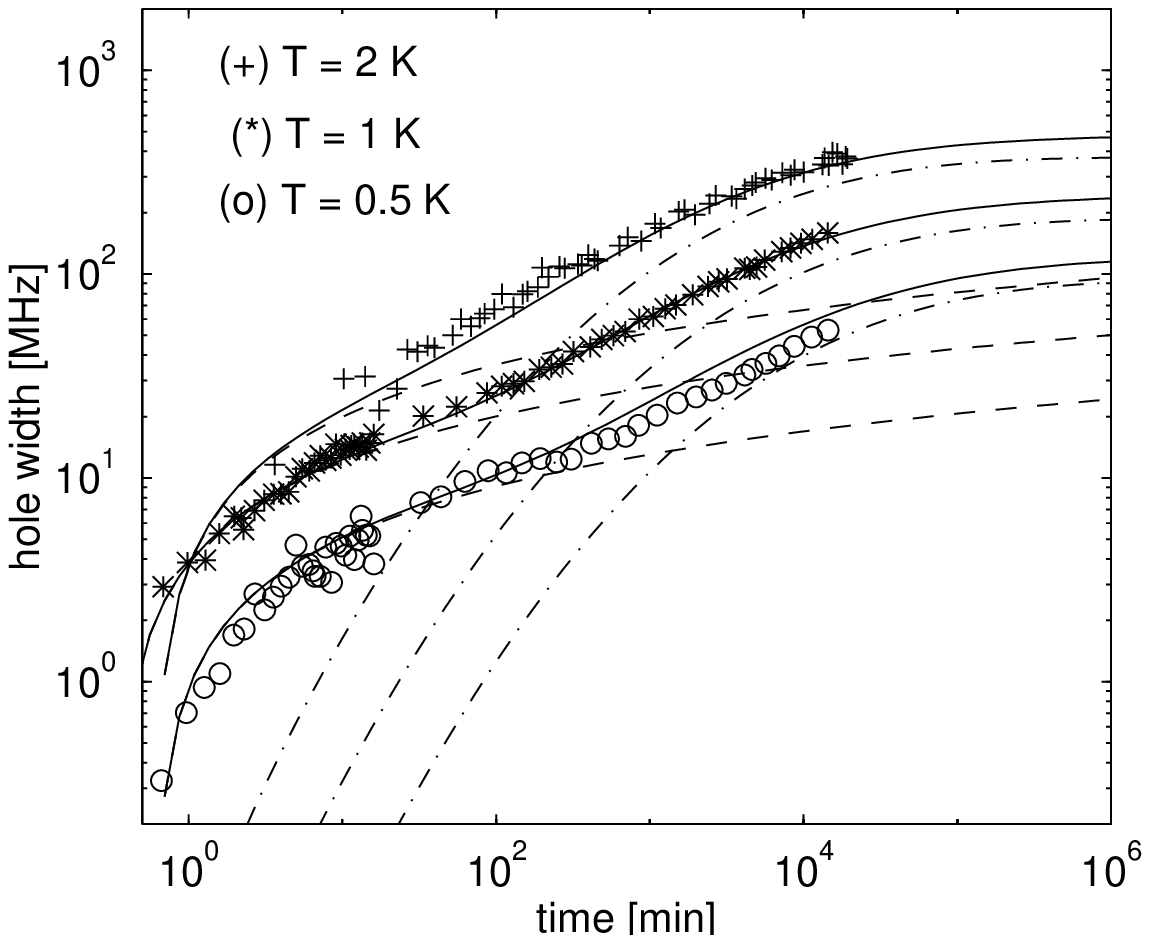}
\end{center}

\noindent FIG. 8: Heuer et al.

\begin{center}
               \epsfig{file=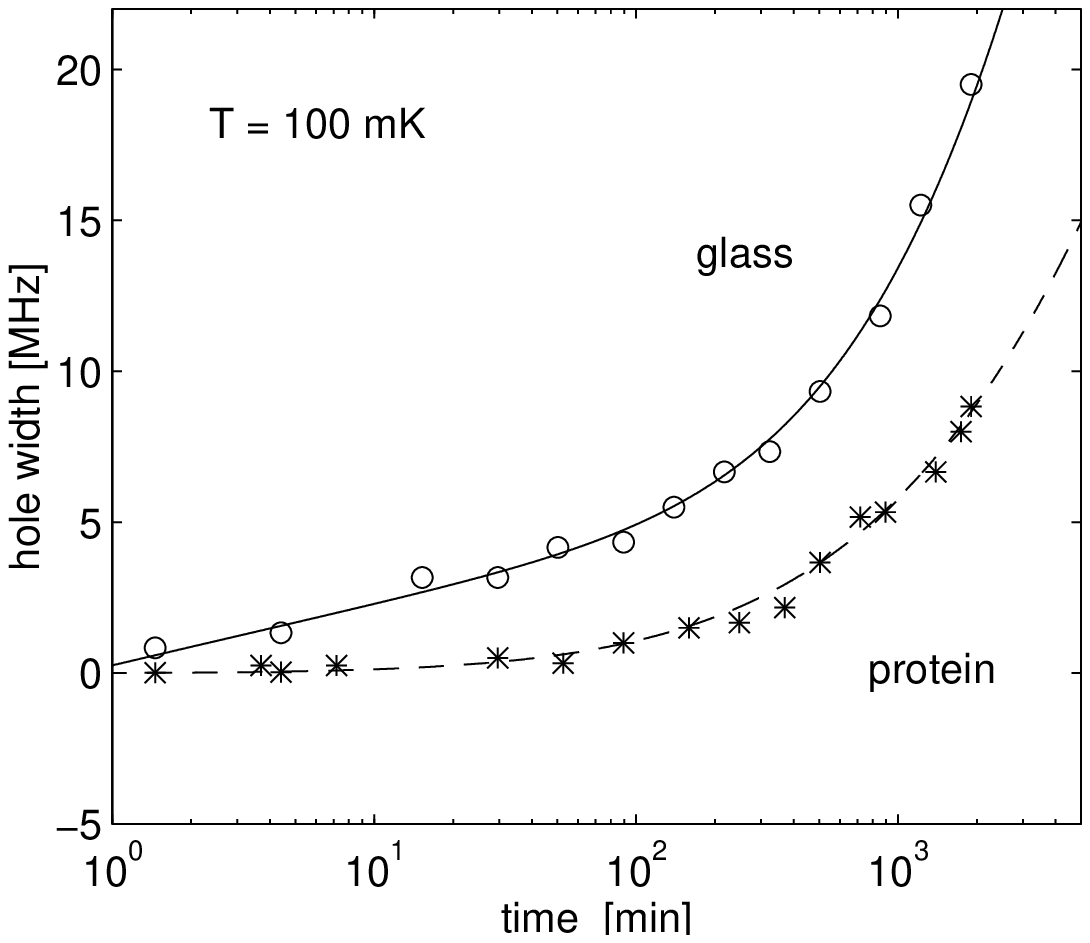}
\end{center}

\newpage

\noindent FIG. 9: Heuer et al.

\begin{center}
               \epsfig{file=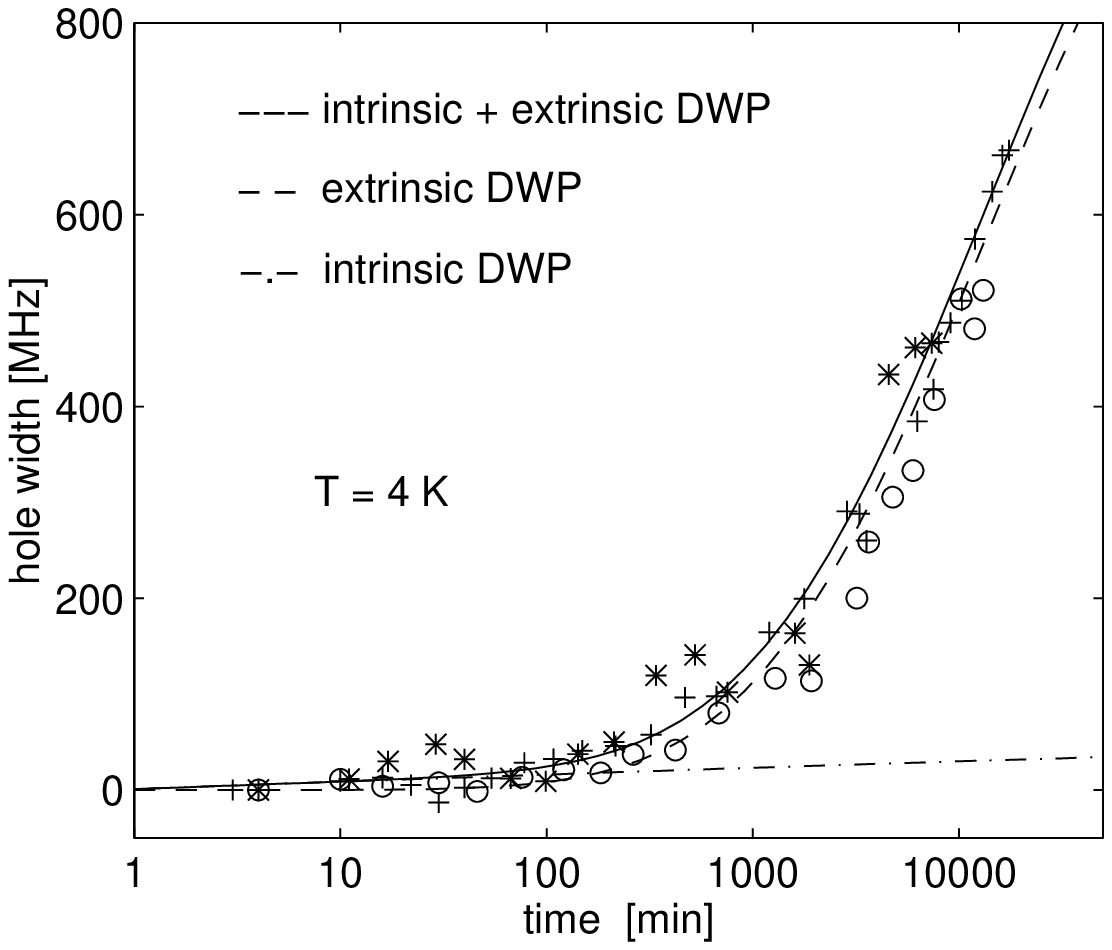}
\end{center}

\begin{references}
\bibitem{cut} Noting that  at  $\omega = 2\pi\, \times \,$15~MHz the peak
temperature  lies at
$T_{\rm max} \approx  12$~K, \cite{FH}  the relation $V_{\rm max} = -k_B T_{\rm
max} \log(\omega\tau_0)$  (cf. Eq. (2.105) in Ref. \CITE{N1})
with $\tau_0 = 10^{-13}$~s provides an estimation
$V_{\rm max}({\rm PMMA}) \approx 220$~K.
Using the same formula to calculate $T_{\rm max}$ at  $\omega = 2\pi\, \times
\,$535~Hz with this value
of $V_{\rm max}$ gives an expected peak temperature of $\sim 10$~K. Comparing
this with the
data of Fig. 12 in Ref. \CITE{Pobell}, one clearly sees that this peak
temperature is too low
to separate the relaxation peak from the plateau.
\end{references}
\end{document}